# Spontaneous microwave platicon frequency microcomb in dispersion-managed microresonators

Wenting Wang[1,2,†,*], Jinkang Lim[1,†], Abhinav Kumar Vinod[1], Mingbin Yu[3], Dim-Lee Kwong[3], and Chee Wei Wong[1,*]

[1] Fang Lu Mesoscopic Optics and Quantum Electronics Laboratory, University of California, Los Angeles, CA 90095, United States of America

[2] Communication and Integrated Photonics Laboratory, Xiongan Institute of Innovation, Chinese Academy of Sciences, Xiong'an New Area, China

[3] Institute of Microelectronics, A*STAR, Singapore 117865, Singapore

[†] These authors contributed equally to this work.

[*] Email: wenting.wang@xii.ac.cn; cheewei.wong@ucla.edu

Temporally stabilized optical pules, confined in microresonators driven by a continuous-wave laser, have attracted tremendous attention due to their fascinating features with many applications. Here we report the observations of mode-locked platicon frequency microcomb formation in normal dispersion dispersion-managed microresonators operating at microwave K-band repetition rate for the first time. Facilitated by the thermally controllable modulated background induced by avoided mode-crossings, various platicon bound state patterns with regular and irregular temporal separation are stably generated due to an additional balance between repulsive and attractive forces resulting from non-trivial interpulse and background electromagnetic field interactions. The number of mode-locked pulses can be switched by forward- and backward-cavity pump detuning and, with increasing pump power, result in stationary bound-state complexes. These experimental observations are in accordance with our nonlinear numerical simulations that includes avoided mode-crossing, anomalous fourth-order dispersion and quality-factor spectral filtering. The observed platicon mode-locked pulses have broad spectral profiles overlapping Kelly-sideband-like parametric oscillation. The single-sideband phase noise of microcomb repetition rate is characterized for the different mode-locked states, comparable with electronic microwave oscillators. The ability to achieve mode-locking in dispersion-managed microresonators provides a platform to reduce pulse timing jitter and enrich the exploration of ultrafast phenomena in microresonators.

# 1. Introduction

Ultrafast nonlinear dynamics in continuous-wave-driven Kerr microresonators have recently been investigated for disciplined ultrafast pulse pattern formation with repetition rates from microwave to terahertz [1-5] and broadband coherent chip-scale frequency comb synthesis [6-8]. The demonstrated frequency microcomb has been applied for massively optical communication [9-11], laser spectroscopy [12-14], precision distance metrology [15,16], coherent terahertz generation [17] and astronomical spectroscopy [18,19]. The excitation of dissipative Kerr soliton (DKS, bright and dark soliton determined by the cavity dispersion) [20,21] benefits from the balance between anomalous dispersion and Kerr nonlinear phase shift, as well as internal dissipation and externally coherent pump driving. The DKS can be spontaneously generated by sweeping the pump laser wavelength across one of the microresonator resonances and anchoring at an effective red pump-cavity detuning with appropriate pump power. Usually the bright DKS undergoes rich cavity dynamics ranging from modulation instability-induced Turing rolls and spatiotemporal chaos [15], to stable cavity solitons pumped at the anomalous dispersion regime. During the soliton formation, the optical Čerenkov radiation [22-24] such as originating from high-order dispersion and Raman-related phenomena in microresonators, can be observed as well [25]. Accessing the stable soliton state experimentally is often hindered due to strong thermal dynamics resulting from thermo-optic effects in microresonators. To mitigate the underlying thermal influences, techniques such as frequency ramping methods including forward- and backward-pump laser scanning [5,26-28] and pump power kicking method [29], pump laser injection locking [30] are proposed and experimentally demonstrated. Turn-key integrated soliton microcombs can be obtained based on the photonic integrated circuits with heterogeneous integration using CMOS foundry materials such as $Si_3N_4$ [31], AlGaAs [32], AlN [33] and CMOS-compatible fabrication processes [34] or edge-coupling technologies combined with a compact laser diode [35,36].

In contrast to the soliton microcomb, the mode-locked microcomb can be formed in the normal dispersion microresonators with different modalities such as dark soliton microcombs [5,21], switching-waves [37] and platicon frequency microcombs [38-40] which exhibit an intrinsically higher optical power conversion efficiency, and larger pulse duty cycle in the time domain. The coherent normal dispersion microcomb can be generated with assistance of quality-factor-dependent spectral filtering [5], mode-coupling such as spatial or polarization mode interaction [41,42], pump laser self-injection locking [43], and external pump laser modulation [44]. Moreover, simultaneously normal-anomalous dispersion microresonators can be fabricated by tuning the waveguide geometry for dispersion engineering to balance the material normal dispersion [45, 46], while further expanding the stability zones.

Here we report the first observations of spontaneous mode-locked platicon frequency microcomb formation with K-band repetition rate in normal dispersion dispersion-managed dissipative microresonators. In analogy to dispersion-managed mode-locked fiber lasers [47], the dispersion-managed dissipative microresonator enhances the mode-locking spectral range [45, 48,49]. The pulses are stabilized with the aid of avoided mode-crossings (AMX) and wavelength-dependent quality-factor spectral filtering. By proper control of the pump laser power and pump-cavity detuning, various mode-locked platicon pulse states are excited and switched. Mode-locking dynamics and transitions related to the platicon pulse are further observed for the first time. The tunability of the primary comb lines is realized by the normal group-velocity dispersion (GVD) and temperature-dependent AMXs. By adjusting pump laser powers, bound states of double-, three-, four-, six-, seven-, and eight-pulses are observed with varying relative temporal separation. These experimental observations are supported by intensity noise characterization, single-sideband phase noise, ultrafast intensity autocorrelation (AC) measurements, and frequency-resolved optical gating (FROG) metrology, along with nonlinear numerical simulations. All measured bound pulses are phase-locked in short- and long-range interactions, clearly manifested by optical spectral interference fringes, single-sideband phase noise, and the intensity AC traces.

## 2. Results

**Figure 1**a shows the scanning electron micrograph of the dispersion-managed $Si_3N_4$ microresonator. The stoichiometric $Si_3N_4$ microresonator features a 7.7 mm circumference, a 1 μm × 0.76 μm width-height cross-section over the bus and curved waveguides to ensure single-mode operation, and seven tapered resonator-waveguide segments of 800 μm lengths each. In each tapered segment, the waveguide width adiabatically increases from 1 to 2.5 μm and vice versa, to finely tune the dispersion and enhance the single-mode mode-locking (detailed in Methods and Supplementary Information Section I). The pulses can exist with the characteristic periodic dispersion oscillation causing pulse stretching and compression behaviors. The numerically evaluated GVD is varied from + 139.3 $fs^2$/mm to -38.7 $fs^2$/mm with the tapered waveguide width from 1 to 2.5 μm. The calculated path-averaged cavity GVD and TOD are +27.9 $fs^2$/mm and -970.8

fs$^3$/mm respectively for a 1590 nm pump wavelength. A commercial near-infrared continuous-wave semiconductor laser pumps the microresonator for the mode-locked pulse generation where a microwave signal is generated through a high-speed photodetector. The pump laser is directly free-space edge-coupled to an engineered inverse taper waveguide on-chip, with coupling loss around 3 dB per facet.

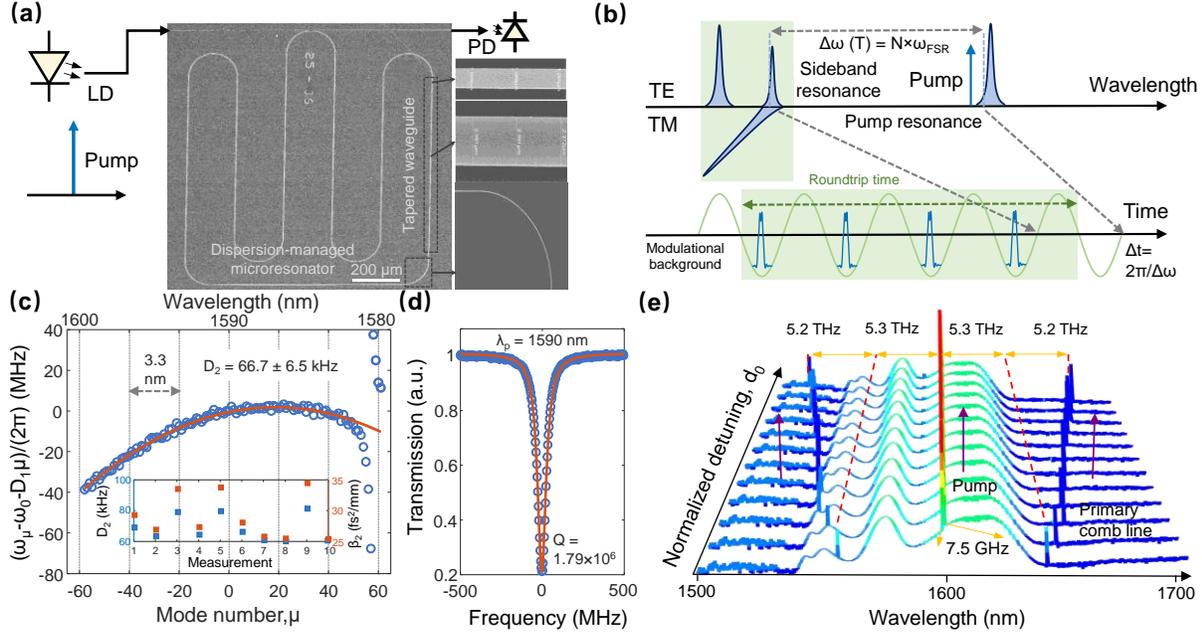

**Figure 1.** Mode-locked platicon frequency microcomb in dispersion-managed microresonators with K-band repetition rate. (a) Schematic of the generation of platicon frequency microcomb in the photonic-chip-based Si$_3$N$_4$ microresonator. The dispersion-managed microresonator constitutes of seven tapered waveguides with varying widths from 1 to 2.5 μm to provide the periodic oscillating group-velocity dispersion. Smaller panels show zoom-in views over the taper waveguides and the curved single-mode spatial mode filter. Scale bar: 200 μm. (b) Conceptual illustration of the spontaneous platicon frequency microcomb formation facilitated by the tunable intracavity modulated field background induced by the avoided mode-crossing (AMX). (c) Measured integrated group velocity dispersion (GVD) of the Si$_3$N$_4$ microresonator with the swept-wavelength interferometry. The measured normal GVD is $\beta_2$ = 28.2 ± 6.4 fs$^2$/mm, obtained from $D_2 = -c/nD_1^2\beta_2$. The FSR ($D_1$) of the device is 19.7 GHz in the microwave K-band. Inset: Multiple measurements of the GVD. (d) Normalized cold-cavity transmission of the single-mode tapered microresonator. The pump resonance at 1590 nm with a loaded $Q$ of 1.79 × 10$^6$. (e) Measured parametric primary comb lines can be continuously tuned – such as over 41 nm in this plot – by controlling the pump-cavity detuning, supported by the AMX or the equivalent anomalous fourth-order dispersion.

**Figure 1**b shows schematic of the mode-locked platicon frequency microcomb generation in the normal dispersion microresonator facilitated by the intracavity field modulated background, triggered by the AMXs. Swept-wavelength interferometry is used to characterize the single-mode cold-cavity dispersion at the fundamental transverse-electric mode (TE$_{00}$) with the TE-polarized pump laser as shown in **Figure 1**c (detailed in Supplementary Information Section II). The microresonator dispersion is determined by analyzing the FSR wavelength dependence with the relation $D_{int}(\mu) = \omega_\mu - \omega_0 - \mu D_1 = D_2\mu^2/2 + D_3\mu^3/6 + \cdots$, where $\omega_\mu$ is the angular frequency of the resonances, $\omega_0$ is the pump laser angular frequency, and $D_3$ is the higher-order dispersion parameter. The measured free spectral range $D_1/2\pi$ = 19.7 GHz with a normal GVD $\beta_2$ of 28.2 ± 6.4 fs$^2$/mm and the corresponding second-order dispersion parameter $D_2/2\pi$ = 66.7 ± 6.5 kHz, based on ten dispersion measurement sets as shown **Figure 1**c inset. The 1590 nm pump resonance is fitted with a Lorentzian lineshape as shown in **Figure 1**d. The coupled resonance linewidth $\kappa/2\pi$ is ≈ 105 MHz, corresponding to the loaded quality factor $Q$ of $1.79 \times 10^6$ (intrinsic $Q$ of $1.86 \times 10^6$). **Figure 1**e shows the widely tunable parametric modulation sidebands when the pump-cavity detuning is continuously swept and controlled.

**Mode-locked platicon frequency microcomb formation**

To deterministically generate the mode-locked pulses, we implement a polarization-assisted resonance pulling method (PARP), which mitigates thermal influences in the microresonator benefiting from the short response time of the polarization modulator (shorter than thermal relaxation time ≈ 1 μs). Instead of accessing the stable soliton step by sweeping pump wavelength, in PARP the pump wavelength is firstly tuned into a close resonance and subsequently an abrupt voltage is applied to the polarization modulator to change the pump laser polarization. This cools down the cavity resonance such that the pump wavelength can access different mode-locking states. To quantify the thermal dynamics in the microresonator, we scan the pump wavelength over the cavity resonance with a 20 nm/s scanning rate to measure the hot cavity transmission as shown in **Figure 2**a. It can be clearly observed that the cavity resonance is red-shifted by the thermo-optic and Kerr nonlinearities with the characteristic discrete step plateau of a negative slope, indicating the AMX after heating the microresonator. The corresponding discrete comb spectral structures are shown in **Figure 2**b which are generated by controlling the pump laser power and pump-resonance detuning. The optical spectrum spans more than 13 THz (-20 dB full-width) and has comb line spacing at the single 19.7 GHz free spectral range of the fundamental mode. The observed spectral asymmetry is attributed to the influence of higher-order dispersion terms. For state III, there are two pairs of strong intensity lines on top of the spectral profile, at 1654.7 nm and 1658.1 nm on the longer wavelength side and at 1525.9 nm and 1528.7 nm on the shorter wavelength side. These are attributable to the tunable primary sidebands. The generation of tunable parametric frequency sidebands can provide a scheme for visible and mid-infrared laser generation. Insets are the correspondingly measured intensity autocorrelation traces that shows the microcomb pulsewidths at, for example, 263 fs and 2 ps.

The pulse-breaking behaviors are verified by an equivalent intracavity average power measurement. In the measurement, a 10-nm bandpass filter centered at 1550 nm is inserted after the mode-locked frequency comb output to select a part of comb (excluding the pump laser), to monitor the equivalent intracavity average power relative to the average power of single mode-locked pulse. An example of the single pulse is shown in **Figure 2**$c_1$ and $c_2$. The average power dynamical evolutions – pulse creation and annihilation – are achieved by forward and backward scanning pump wavelength. First, we set the pump power at 2W and tune the pump wavelength forward. The pulse formation (i.e. pulse breaking) is subsequently observed with the power jump shown in **Figure 2**$c_1$. Secondly, we increase the pump power to 2.15W and obtain multiple mode-locked pulses. Subsequently we tuned the pump wavelength backward to observe the pulse annihilation and likewise stably access the single mode-locked pulses. The monitored equivalent intracavity power is suggestive of dynamical pulse breakup and pulse annihilation to form different bound-states.

**Figure 2**d shows the spectral evolution with the scanned pump wavelength centered at 1590.11 nm and over 108× the cavity linewidth ($\approx$ 11.3 GHz). Reducing the pump-cavity detuning results in the intracavity power accumulation leading to the primary comb line (I), single mode-locked pulse (II) and mode-locked paired pulses (III). The tunable phase-matched sidebands support the evidence of the positive GVD and temperature-dependent AMXs in the hot microresonator [50]. The AMX provides an extra phase shift term to compensate Kerr nonlinear phase shift and normal GVD for mode-locked pulse formation, akin to higher-order dispersion [51]. **Figure 2**e shows the RF intensity noise of the mode-locked comb (III) by injecting a selected 10-nm span of the comb into a high-speed photodetector. We examined the RF intensity noise up to 1 GHz, much larger

than the cavity linewidth. The comb state has a low intensity noise at the photodetector background noise floor, with a 19.7 GHz repetition rate beat note measured by a fast 20 GHz InGaAs photodiode. The RF beat is narrow linewidth at 200 Hz without additional linewidth broadening and multiple beat notes such as breathing tones. Then the center wavelength tunability of the generated mode-locked frequency comb is examined by changing the pump from 1589 to 1591 nm as illustrated in **Figure 2**f. We observe the platicon parametric sidebands tunability which is related to the effective pump-resonance detuning.

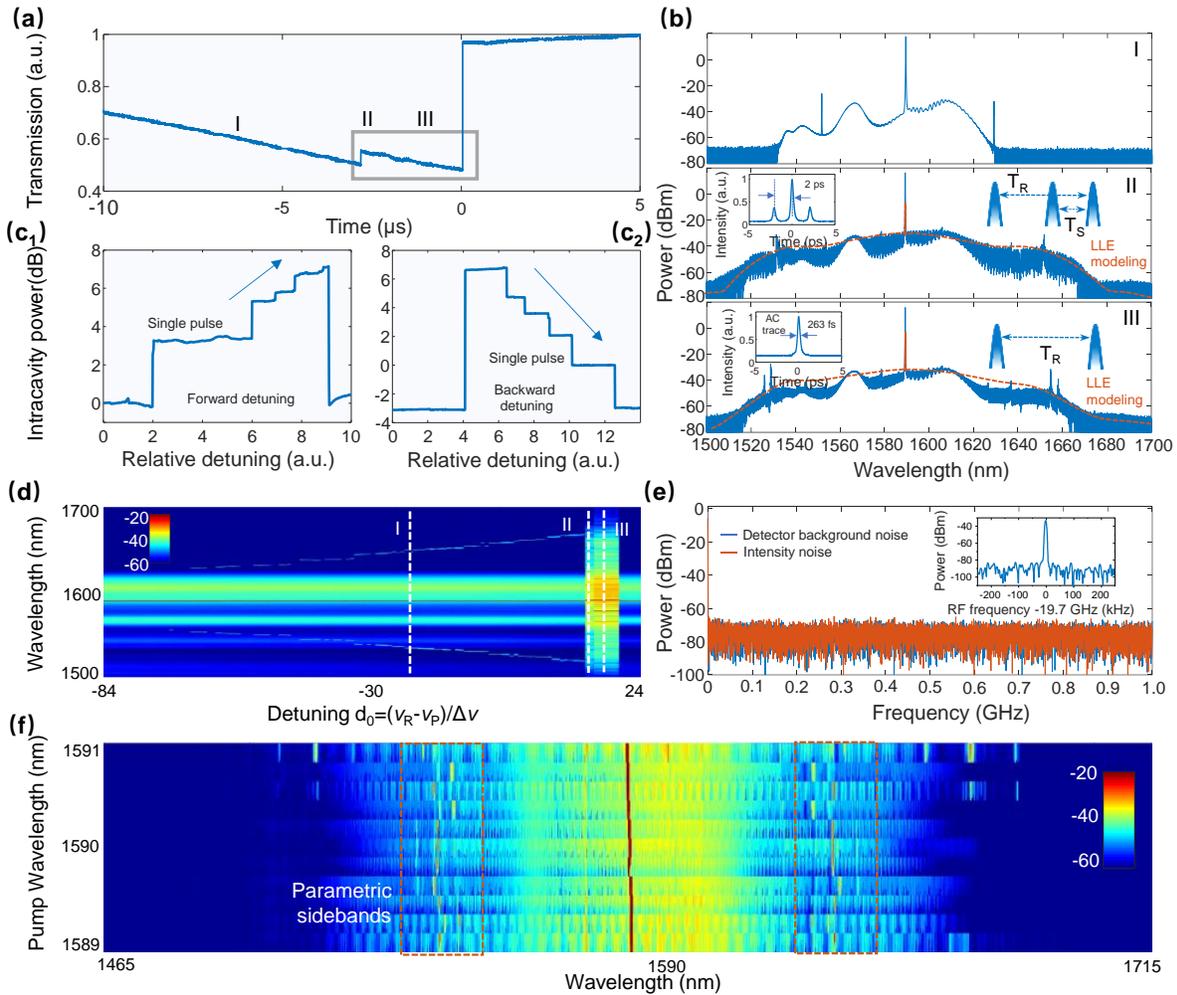

**Figure 2.** Mode-locked platicon frequency microcomb formation. (a) Measured transmission when the pump wavelength is scanned at 20 nm/s over the cavity resonance, with the characteristic triangular thermal dragging response and soliton discrete step. (b) Controlled frequency comb spectra under three different detunings, from top panel to bottom: primary comb lines, double mode-locked pulses, and single mode-locked pulse overlapping with Lugiato-Lefever equation (LLE) modelled frequency microcomb. Inset of II shows the intensity autocorrelation trace of the double-pulse. Inset of III shows an example autocorrelation-measured 263 fs single pulse pulsewidth. ($c_1$) and ($c_2$) Intracavity power evolution of microresonator when the pump wavelength is tuned forward and backward, with power relative to the single mode-locked pulse. The multiple discrete step plateau is suggestive of dynamical pulse breakup with forward detuning and pulse annihilation with backward detuning. (d) Dynamical evolution with pump-cavity detuning, illustrating transitions from tunable primary comb lines (I) to a double mode-locked pulse state (II) and subsequently to the single mode-locked pulse state (III). (e) RF intensity noise of the mode-locked pulse (II and III) measured up to 1 GHz, plotted along with the photodetector background noise, indicating the low phase noise operation. Inset shows the fundamental RF linewidth of the mode-locked pulse measured with a 10 Hz resolution bandwidth. (f) Mode-locked platicon frequency comb tunability. The frequency comb was initially generated with the pump wavelength at 1590 nm. Subsequently the pump is tuned from 1589 to 1591 nm.

**Temporal characterization of the mode-locked platicon pulse**

To shed light on the temporal property of the generated mode-locked pulses, we performed the second-harmonic non-collinear intensity AC and sub-femtojoule sensitivity FROG measurements [5] (detailed in Methods). In contrast to a single pulse, the platicon bound states comprise of multiple pulses with relative phases $\varphi$ separated by a time $\tau$ and can be resolved by the spectral offset in the interference fringes relative to spectral envelope $\delta v$ with a relation of $\varphi = 2\pi\delta v/\Delta v$. $\Delta v$ (=$1/\tau$) is determined by the pulse temporal separation. **Figure 3**a-h show the characteristic optical spectra. The single pulse, shown in the top row, has a smooth spectral profile while the bound pulses have periodically modulated spectral profiles due to pulse interference. The modulation period and depth are represented in the optical spectra of the bound pulses. The modulation depth is related to the phase coherence between the individually bound pulses. Asymmetry of the optical spectra is caused mainly by the microresonator TOD. At the spectral edges, we observe clear high-intensity continuous wave components, which are the tunable parametric sidebands [52]. All of the comb spectra show a comb line spacing of one FSR. **Figure 3**e-h show the corresponding intensity AC traces, indicating the firmly mode-locked behaviors and pulse temporal separation. The measured pulse duration is 263 fs corresponding to the bandpass filtered 40 nm optical spectral bandwidth with an extra pulse chirp.

Benefiting from pump laser suppression, the measured AC traces have high peak-to-background contrast. The bound states repetition is separated by the 50.7 ps microresonator round-trip time, corresponding to the comb 19.7 GHz FSR. The AC traces show the relative intensity ratios between adjacent pulses of the bound states. With the AC number of peaks $M$ (= $2N - 1$, where $N$ as the number of pulses), an intensity ratio of 1:2:1 in the AC trace would indicate two pulses of the same energy. We also observed regular and irregular bound states, with equal and unequal temporal separations respectively. All of the measured AC traces show the additional weak intensity peaks indicating the non-perfect-crystal property of the pulse patterns. Insets further illustrate the measured wavelength-time 2D FROG spectrograms with a 16 ps time delay.

The bound pulses – stationary multiple pulses in a cavity – stem from the balance between attractive and repulsive forces. The balanced linear and nonlinear interaction is mediated by parametric gain, loss, Kerr nonlinearity and AMX contributing to the same attractor. Experimentally the fully modulated spectral fringes are evidence of tight phase-locking and a constant phase relationship in the twin-pulses. With the controlled pump power and detuning in this dissipative dispersion-managed stretched cavity, more complex bound patterns can be formed, with and without regular temporal spacing. **Figures 3**i-l illustrate cases where four, six, seven and eight pulses are tightly phase-locked respectively. Some have equal pulse temporal separation and energies in single pulse bunches while others have unequal separations. Moreover, the bound pulse can function as a unit binding with another bound pulse unit as shown in **Figure 3**l. In this double bound state case of eight pulses, the relation between intensity peak number $M$ and pulse number $N$ is $M = 2N - 2$ when pulses distribute with a periodic order. We note this eight-pulse double irregular bound state has two pulse bunches with regular temporal separations respectively. With assistance of second-harmonic-based FROG spectroscopy and intensity autocorrelation, we observed the steady-state spatio-temporal property of the bound states of cavity soliton in the dispersion-managed stretched cavity microresonator.

By numerically solving the Lugiato-Lefever equation (LLE) with parameters close to the dispersion-managed microresonator used in our measurements, we examined the microresonator with positive GVD, avoided mode-crossings, and wavelength-dependent quality-factor spectral filtering effect to elucidate the mode-locking mechanism. We carried out the numerical simulation with 4,001 optical resonances around the pumping mode for the platicon single- and multiple pulses formation (detailed in Methods and Supplementary Information Section III). By sweeping the pump-cavity detuning, good agreement in optical spectra is observed between the modeling and experimental results with examples shown in **Figures 3**a and d. We consider these dual pulses as bound states where the bound pulses are linked by extended periodically oscillatory tails triggered by the polarization mode-crossing and with little net drift of bound pulses from the low phase noise characteristics to be detailed in the next section.

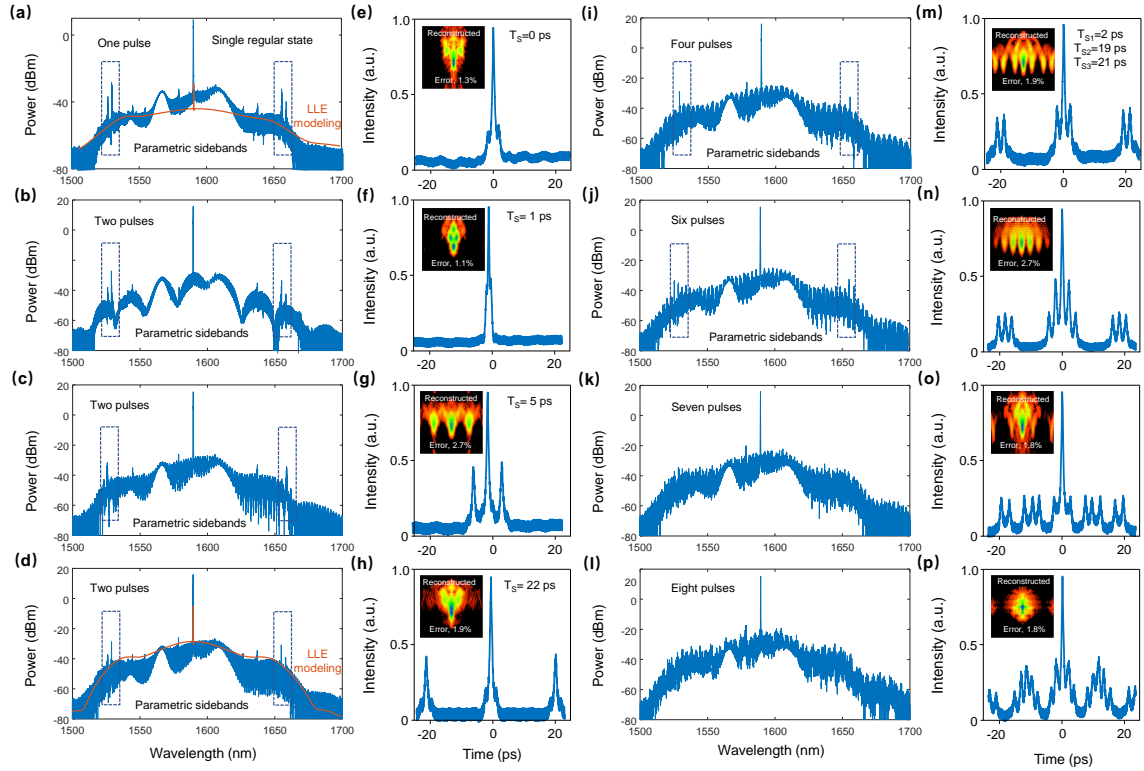

**Figure 3.** Mode-locked platicon bound state pulse patterns with ultrafast temporal characterization. (a)-(d) and (i)-(l) Optical spectra of the single, double, triple, quadruple, sextuple, septuple, and octuple pulses respectively with the marked parametric sidebands which are similar with the resonant sideband in fiber mode-locked lasers [48]. The single pulse is shown in the top row for comparison. The modulated spectrally profiles arise from the bound pulses spectral interference where the modulation frequency and depth are determined by the temporal separation and relative phase difference between the bound pulses. The pulse patterns exhibit a single FSR comb line spacing since they are not perfect crystal states. (e)-(h) and (m)-(p) Intensity autocorrelation (AC) traces of the corresponding pulses over half microresonator round-trip delay time for the selected microcomb spectra without the pump line. The coherence of the bound pulses is verified by the AC traces and single-sideband phase noise measurement, which illuminates the pulse temporal separation and distribution. Insets: reconstructed 2D frequency-resolved optical gating (FROG) spectrograms, with reconstruction via genetic algorithms for the phase retrieval.

**Repetition-rate single-sideband phase noise characterization of the mode-locked platicon pulses.**

The schematic of the mode-locked pulse generation and characterization setup is shown in **Figure 4**a which comprises of pump laser, microresonator followed by an oscilloscope to record the optical transmission, an optical spectrum analyzer for recording optical spectra, an RF electrical spectrum analyzer for monitoring intensity noise and repetition rate RF beat note, a second-harmonic intensity autocorrelator and FROG spectroscopy system, and a signal source analyzer. The total fiber-chip-fiber transmission loss of the microresonator, including propagation and coupling losses, is less than 6 dB. A custom-built temperature-controlled stage mounts the chip with 100 mK temperature stability over one hour. The polarization beam splitter (PBS) selects the TE-polarized lightwave and introduces a polarization-dependent power change. An important property of this microcomb is that the 19.7 GHz repetition rate is directly detectable with our high-speed photodetector and signal source analyzer, enabling the cross-correlation phase noise examination of the microcomb repetition rate at K-band. **Figure 4**b shows the measured single-sideband phase noise at different mode-locked states, including the corresponding RF beat notes after frequency division as shown in insets. For the RF signal spectrum and phase noise measurement, a 10 nm bandwidth optical filter (1560 to 1570 nm) is used to avoid power saturation from the pump laser. The selected optical spectra are amplified with the same power amplification for all mode-locked states for proper comparison. A 20 GHz InGaAs photodiode is utilized to detect the platicon repetition rate. The RF FSR beat notes show a clean RF signal over a 5 MHz spectral range with frequency of 19.7/8 GHz and signal-to-noise ratio of more than 70 dB. This confirms the phase-locking behavior and is comparable to current high-performance K-band microwave oscillators.

Without any external feedback control to stabilize the mode-locked microcombs, the phase noise of the repetition rate close to 19.7 GHz is ≈ -90 dBc/Hz at 10 kHz Fourier offset frequency. The phase noise spectra exhibit a power-law slope of 30 dB/decade from 10 Hz to 40 kHz which is mainly caused by the free-running pump-resonance detuning instability [53,54]. The noise peaks from 10 kHz to 70 kHz result from the intensity noise of the high optical power amplifier. The light grey curve is the measured relative intensity noise after accessing the microresonator resonance. The measured phase noise spectra are related to the pump laser noise. Other chip-scale photonics-based microwave oscillators with similar frequency are included in **Figure 4**b for comparison. For our single-pulse platicon microcomb, the measured phase noise has better phase noise than the other multi-pulse bound states at lower offset frequency below than 10 kHz. The noise degradation is attributed to the spectral interference and pump-resonance detuning noise. Above this frequency, the preamplifier spontaneous emission will be the dominant limit at -116 dBc/Hz for 1 MHz Fourier offset frequency, which is related to its incident optical power. An extra 6 dB noise degradation is observed over the different mode-locked states. We can anticipate the phase noise improvement in a fully packaged microcomb and with optimized feedback control.

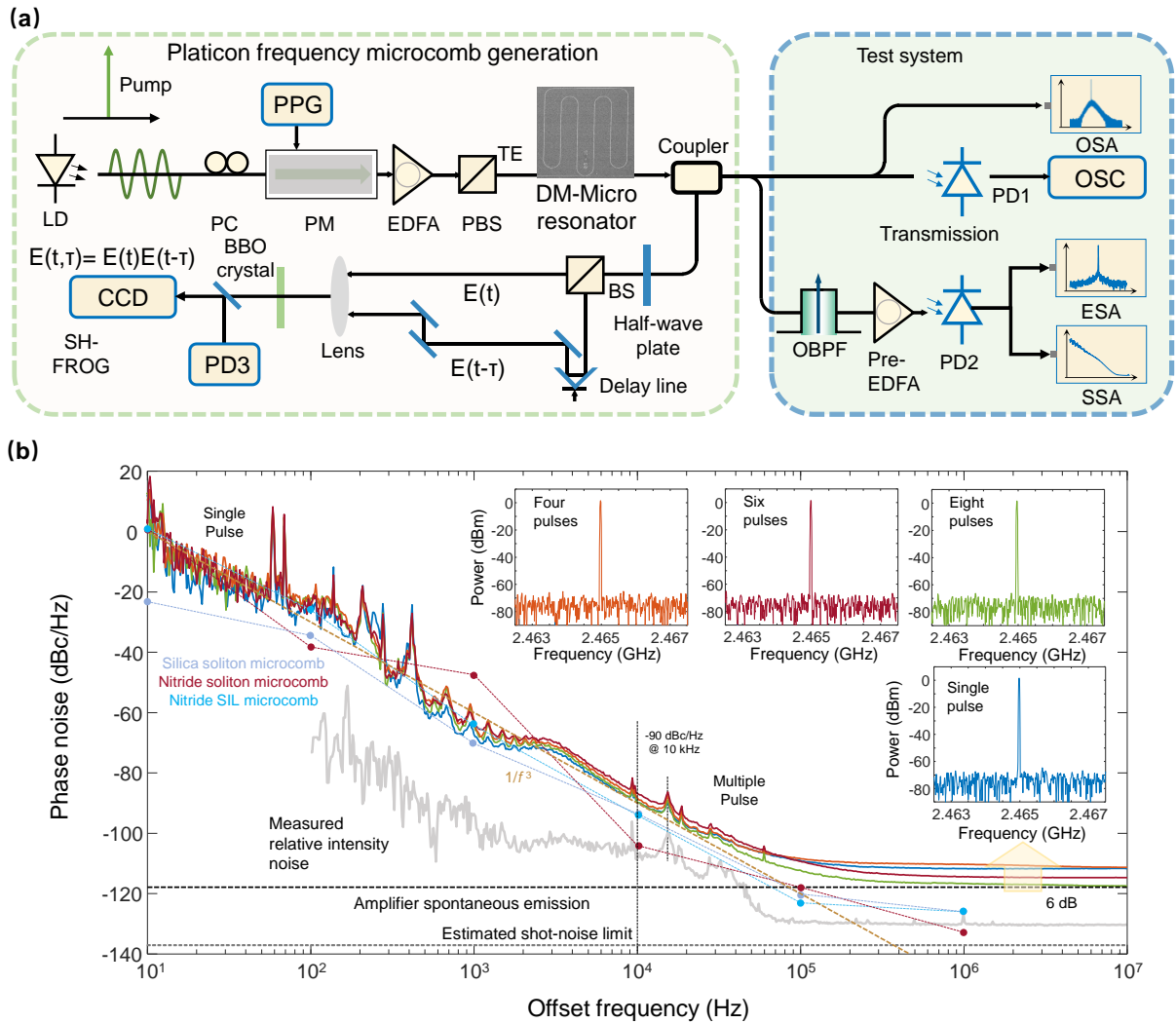

**Figure 4.** Repetition-rate single-sideband phase noise characterization of the mode-locked pulse patterns. (a) Experimental setup to simultaneously monitor optical transmission with characteristic thermal triangle, the optical spectra, the RF spectra including intensity noise and beat notes between adjacent frequency comb modes, the second-order non-collinear intensity autocorrelation

and FROG spectrograms, and a signal source analyzer. LD: laser diode; PC: polarization controller; PM: polarization modulator; PPG: pulse pattern generator; EDFA: erbium-doped fiber amplifier; PBS: polarization beam splitter; OSA: optical spectrum analyzer; OSC: oscilloscope; ESA: electrical spectrum analyzer; SSA: signal source analyzer; PD: photodetector; OBPF: optical bandpass filter; BS: beam splitter; BBO: β-barium borate; and CCD: charge-coupled device. (b) Single-sideband phase noise spectra of the 19.7 GHz repetition rate beat notes for the different mode-locked states. The measured relative intensity noise from pump laser and pump amplifier is also shown in grey curve, supporting the coherent low-noise regime. The intensity noise peaks are reproduced in the repetition rate phase noise indicating the existence of the intensity-to-phase noise conversion. An external comparison of the phase noises of the (silica and silicon nitride) soliton [26,31] and self-injection-locked (SIL) microcombs [43] are also included with the dot-dashed lines, normalized to the 19.7 GHz repetition rate. Insets: the corresponding repetition-rate beat note spectra after 8 times frequency division.

## 3. Discussion

Mode-locked platicon femtosecond pulses initiated by the tunable parametric pattern and stabilized by avoided mode-crossing and quality-factor-related spectral filters in a $Si_3N_4$ dispersion-managed microresonator are observed for the first time. The microresonator features positive GVD which is distinct from microresonators with negative GVD for the typical bright soliton formation. With the measured broad optical spectra, the dependence of the tunable parametric sidebands on the pump-cavity detuning is observed in the microresonator, which provides a signature for the spontaneous mode-locked pulse generation. At higher pump powers, bound states of the platicon pulses with different pulse temporal separations, different pulse numbers $N$ and peak numbers $M$ are experimentally demonstrated for the first time. The stationary bound pulses with invariant interpulse temporal displacement and constant phase difference are stabilized via the balanced attractive and repulsive forces, mediated by the AMX induced intracavity field modulated background in the microresonator. The temporal separations are sensitive to the initial pump conditions and microresonator parameters, while the number of pulses of each bound state depends on the pump powers and pump-cavity detuning. All mode-locked pulses are characterized by the optical spectra, intensity AC traces, FROG spectrograms, RF intensity noise, and single-sideband phase noise. The experimental observations of the mode-locked platicon pulses not only provides a new route for mode-locking but also offer a platform of choice via the platicon microcomb bound states for many applications.

## 4. Experimental Section/Methods

*Device fabrication and dispersion characterization:* The fabrication procedure of the dispersion managed microresonator starts with a 3 μm thick $SiO_2$ layer, first deposited via plasma-enhanced chemical vapor deposition (PECVD) on *p*-type 8-inch silicon wafers to serve as the under-cladding oxide. A 760 nm nitride layer is then deposited onto the top of the under-cladding for ring resonators fabrication. The resulting nitride layer is patterned by optimized 248 nm deep-ultraviolet lithography and etched down to the buried oxide cladding via optimized reactive ion dry etching. The etched sidewalls have an etch verticality of 88 degrees characterized by SEM. The nitride rings are then over-cladded with a 3 μm thick oxide layer, deposited initially with LPCVD for 0.5 μm and then with PECVD for 2.5 μm. The Santec TSL-510 laser is swept through its full wavelength range of 100 nm at 100 nm/s tuning speed with a 1 mW output power and 1% of the laser output is directed into a fiber coupled hydrogen cyanide gas cell (HCN-13-100, Wavelength References Inc.) for the absolute wavelength calibration. The microresonator and gas cell transmission are recorded during the laser sweeping by a data acquisition system with sampling resolution 1.6 MHz derived from an unbalanced fiber Mach-Zehnder Interferometer (MZI). The measured results are in the supplementary information II **Figure S**4.

*Numerical simulations:* Taking the avoided mode-crossing or fourth-order dispersion (FOD) into consideration and wavelength-dependent quality-factor spectral filtering effects, the platicon mode-locked pulse formation dynamic is numerically modeled with Lugiato-Lefever equation (LLE) written as:

$$T_R \frac{\partial E(T,t)}{\partial T} + i \left( \frac{\beta_{2\Sigma}}{2} \frac{\partial^2}{\partial t^2} - \gamma |E|^2 \right) E(T,t) = -\left( \alpha + \frac{T_c}{2} + id_0 \right) E(T,t) + \sqrt{T_c} E_{in},$$

where cavity round-trip time $T_R$ = 50.7 ps, $\beta_{2\Sigma}$ = 21.5 fs$^2$/mm, $\beta_{4\Sigma}$ = -440 fs$^4$/mm, effective nonlinearity coefficient $\gamma = (\omega/c) \times (n_2/A_{eff}) \approx 1$ W$^{-1}$m$^{-1}$, $n_2 \approx 2.5 \times 10^{-19}$ m$^2$/W, calculated effective mode area is $A_{eff} \approx 1$ μm$^2$, $\alpha$ = 0.016, $T_c$ = 0.009. Other parameters are the same value with the measured ones. The pump power and wavelength were set at $P_{in}$ = 2 W and $\lambda_p$ = 1590 nm. We numerically solved the equation with split-step Fourier method starting from the quantum noise. Mode frequency shift is incorporated in the simulation through the empirical two-parameters model: $a/2/(\mu - b - 0.5)$, where *a* is the max modal frequency shift, *μ* and *b* are mode number and mode number for max modal frequency shift, respectively. Multiple mode frequency shifts are included to obtain the stable mode-locked pulse states. Higher-order Gaussian filters are included in the simulation as well. 4,001 modes centered at the pump are incorporated in the LLE simulation running over $1 \times 10^5$ roundtrips until the solution reaches steady-state. By proper setting the pump power and pump wavelength sweeping speed, we can numerically simulate the single pulse and double pulses formation with different temporal separation corresponding to the experiments.

*Second-harmonic-based intensity autocorrelation and frequency-resolved optical gating spectroscopy:* A short-pass optical filter selects a 40 nm optical spectrum from 1530 to 1570 nm and suppresses the pump laser by 30 dB. A C-band optical amplifier (Amonics, AEDFA-PA-35-B-FA) amplifies the average power of the selected optical signal to 20 mW for AC trace and FROG spectrogram measurements. Careful dispersion compensation is conducted with dispersion compensation fibers. The amplified and dispersion-compensated optical signal is injected into a β-barium-borate-crystal-based interferometer for second harmonic generation incorporating with a femtojoule-sensitivity silicon photodetector (Thorlabs, PDF10A) for autocorrelation measurements and a cryogenically-cooled deep-depletion 1024×256 Si CCD array-based grating spectrometer (Horiba Jobin Yvon CCD-1024×256-BIDD-1LS) for FROG spectrogram measurements. The spectrometer is cooled down to -120°C and the center wavelength set at 775 nm with 20 nm optical spectral bandwidth. The exposure time is 1 s and the temporal delay time is set at 16 ps. The measured FROG spectrograms have a size of 301×1024 pixels which can be reconstructed by genetic algorithm, converging to the retrieved temporal and spectral fields. The retrieval errors are less than 3%. The difference between measured $S_{me}(\omega,t)$ and reconstructed $S_{re}(\omega,t)$ spectrograms are quantitatively defined by the relation of $\varepsilon = \sqrt{\frac{1}{N^2}\sum_{i,j=1}^{N}|S_{me}(\omega,t) - S_{re}(\omega,t)|^2}$, where $N$ is the iteration number of the reconstruction.

**Supporting Information**

Supporting Information is available from the Wiley Online Library or from the author.

**Acknowledgements**

The authors acknowledge fruitful discussions with Prof. Heng Zhou on simulations, Hyunpil Boo on the scanning electron micrograph of the microresonator, Jinghui Yang and Shu-Wei Huang for laying out the microresonator. We also thank general discussions with Yoon-Soo Jang, Xinghe Jiang, Tristan Melton, Allen Chu, Alwaleed Aldhafeeri, and Alex Wenxu Gu. We acknowledge financial support from the the Office of Naval Research (N00014-21-1-2259), the Lawrence

**Author contributions**

W.W. and C.W.W. initiated the project. W.W. conducted the experiments. W.W. and J.L. analyzed the data and performed the simulations. J.Y. designed the microresonator. A.K.V. and J.Y. contributed to the experiments. M.Y. and D.-L.K. performed the device nanofabrication. W.W., J.L., and C.W.W. contributed to writing and revision of the manuscript.

**Conflict of interest**

The authors declare no competing financial interests.

**References**

1. T. J. Kippenberg, A. L. Gaeta, M. Lipson, and M. L. Gorodetsky, Dissipative Kerr solitons in optical microresonators. *Science* **361**, 8083 (2018).
2. B. C. Yao, S.-W. Huang, Y. Liu, A. K. Vinod, C. Choi, M. Hoff, Y. N. Li, M. B. Yu, Z. Y. Feng, D. L. Kwong, Y. Huang, Y. J. Rao, X. F. Duan, and C. W. Wong, Gate-tunable frequency combs in graphene-nitride microresonators. *Nature* **558**, 410-415 (2018).
3. B. Stern, X. C. Ji, Y. Okawachi, A. L. Gaeta, and M. Lipson, Battery-operated integrated frequency comb generator. *Nature* **562**, 401-405 (2018).
4. N. G. Pavlov, S. Koptyaev, G. V. Lihachev, A. S. Voloshin, A. S. Gorodnitskiy, M. V. Ryabko, S. V. Polonsky, and M. L. Gorodetsky, Narrow-linewidth lasing and soliton Kerr microcombs with ordinary laser diodes. *Nat. Photon* **12**, 694–698 (2018).
5. S.-W. Huang, J. F. McMillan, J. Yang, A. Matsko, H. Zhou, M. Yu, D.-L. Kwong, L. Maleki, and C. W. Wong, Mode-locked ultrashort pulse generation from on-chip normal dispersion microresonators, *Phys. Rev. Lett.* **114**, 053901 (2015).
6. P. Del Haye, A. Schliesser, O. Arcizet, T. Wilken, R. Holzwarth, and T. J. Kippenberg, Optical frequency comb generation from a monolithic microresonator. *Nature* **450**, 1214-1217 (2007).
7. D. T. Spencer, T. Drake, T. C. Briles, J. Stone, L. C. Sinclair, C. Fredrick, Q. Li, D. Westly, B. R. Ilic, A. Bluestone, N. Volet, T. Komljenovic, L. Chang, S. H. Lee, D. Y. Oh, M. G. Suh, K. Y. Yang, M. P. Pfeiffer, T. J. Kippenberg, E. Norberg, L. Theogarajan, K. Vahala, N. R.

Newbury, K. Srinivasan, J. E. Bowers, S. A. Diddams, and S. B. Papp, An optical-frequency synthesizer using integrated photonics. *Nature* **557**, 81-84 (2018).

8. S. W. Huang, J. H. Yang, M. Yu, B. H. McGuyer, D. L. Kwong, T. Zelevinsky, and C. W. Wong, A broadband chip-scale optical frequency synthesizer at $2.7\times10^{-16}$ relative inaccuracy. *Sci. Adv.* **2**, e1501489 (2016).

9. P. M. Palomo, J. N. Kemal, M. Karpov, A. Kordts, J. Pfeifle, M. H. P. Pfeiffer, P. Trocha, S. Wolf, V. Brasch, M. H. Anderson, R. Rosenberger, K. Vijayan, W. Freude, T. J. Kippenberg, and C. Koos, Microresonator-based solitons for massively parallel coherent optical communications. *Nature* **546**, 574-578 (2017).

10. J. Pfeifle, V. Brasch, M. Lauermann Y. M. Yu, D. Wegner, T. Herr, K. Hartinger, P. Schindler, J. S. Li, D. Hillerkuss, R. Schmogrow, C. Weimann, R. Holzwarth, W. Freude, J. Leuthold, T. J. Kippenberg, and C. Koos, Coherent terabit communications with microresonator Kerr frequency combs. *Nat. Photon* **8**, 375–380 (2014).

11. A. Fülöp, M. Mazur, A. Lorences-Riesgo, Ó. B. Helgason, P. H. Wang, Y. Xuan, D. E. Leaird, M. H. Qi, and P. A. Andrekson, A. M. Weiner, and V. T. Company, High-order coherent communications using mode-locked dark-pulse Kerr combs from microresonators. *Nature Commun.* **9**, 1598 (2018).

12. M. G. Suh, Q. F. Yang, K. Y. Yang, and K. J. Vahala, Microresonator soliton dual-comb spectroscopy. *Science* **354**, 600-603 (2016).

13. A. Dutt, C. Joshi, X. C. Ji, J. Cardenas, Y. Okawachi, K. Luke, A. L. Gaeta, and M. Lipson, On-chip dual-comb source for spectroscopy. *Sci. Adv.* **2**, e1701858 (2018).

14. M. J. Yu, Y. Okawachi, A. G. Griffith, N. Picqué, M. Lipson, and A. L. Gaeta, Soliton-chip-based mid-infrared dual-comb spectroscopy. *Nature Commun.* **9**, 1869 (2018).

15. M. G. Suh and K. J. Vahala, Soliton microcomb range measurement. *Science* **359**, 884-887 (2018).

16. P. Trocha, M. Karpov, D. Ganin, M. H. P. Pfeiffer, A. Kordts, S. Wolf, J. Krockenberger, P. M. Palomo, C. Weimann, S. Randel, W. Freude, T. J. Kippenberg, and C. Koos, Ultrafast optical ranging using microresonator soliton frequency combs. *Science* **359**, 884-887 (2018).

17. S.-W. Huang, J. Yang, S.-H. Yang, M. Yu, D.-L. Kwong, T. Zelevinsky, M. Jarrahi, and C. W. Wong, Globally stable microresonator Turing pattern formation for coherent high-power THz radiation on-chip. *Phys. Rev. X* **7**, 041002 (2017).

18. E. Obrzud, M. Rainer, A. Harutyunyan, M. H. Anderson, M. Geiselmann, B. Chazelas, S. Kundermann, S. Lecomte, M. Cecconi, A. Ghedina, E. Molinari, F. Pepe, F. Wildi, F. Bouchy, T. J. Kippenberg, and T. Herr, A microphotonic astrocomb. *Nat. Photon* **13**, 31–35 (2019).
19. M. G. Suh, X. Yi, Y. H. Lai, S. Leifer, I. S. Grudinin, G. Vasisht, E. C. Martin, M. P. Fitzgerald, G. Doppmann, J. Wang, D. Mawet, S. B. Papp, S. A. Diddams, C. Beichman, and K. J. Vahala, Searching for exoplanets using a microresonator astrocomb. *Nat. Photon* **13**, 31–35 (2019).
20. T. Herr, V. Brasch, J. D. Jost, C. Y. Wang, N. M. Kondratiev, M. L. Gorodetsky, and T. J. Kippenberg, Temporal solitons in optical microresonators. *Nat. Photon* **8**, 145–152 (2014).
21. X. X. Xue, Y. Xuan, Y. Liu, P. H. Wang, S. Chen, J. Wang, D. E. Leaird, M. H. Qi, and A. M. Weiner, Mode-locked dark pulse Kerr combs in normal-dispersion microresonators. *Nat. Photon* **9**, 594–600 (2015).
22. H. Zhou, S.-W. Huang, X. Li, J. F. McMillan, C. Zhang, K. K. Y. Wong, M. Yu, G.-Q. Lo, D.-L. Wong, K. Qiu, and C. W. Wong, Real-time dynamics and cross-correlation gating spectroscopy of free-carrier Drude solitons, *Light: Science & Applications* **6**, e17008 (2017).
23. V. Brasch, M. Geiselmann, T. Herr, G. Lihachev, M. H. P. Pfeiffer, M. L. Gorodetsky, and T. J. Kippenberg, Photonic chip–based optical frequency comb using soliton Cherenkov radiation. *Science* **351**, 357–360 (2016).
24. Q. Li, T. C. Briles, D. A. Westly, T. E. Drake, J. R. Stone, B. R. Ilic, S. A. Diddams, S. B. Papp, and K. Srinivasan, Stably accessing octave-spanning microresonator frequency combs in the soliton regime. *Optica* **4**, 193-203 (2017).
25. Q. F. Yang, X. Yi, K. Y. Yang, and K. Vahala, Stokes solitons in optical microcavities. *Nat. Phys.* **13**, 53–58 (2016).
26. X. Yi, Q.-F. Yang, K. Y. Yang, M.-G. Suh, and K. Vahala, Soliton frequency comb at microwave rates in a high-*Q* silica microresonator. *Optica* **2**, 1078-1085 (2015).
27. H. Guo, M. Karpov, E. Lucas, A. Kordts, M. H. P. Pfeiffer, V. Brasch, G. Lihachev, V. E. Lobanov, M. L. Gorodetsky, and T. J. Kippenberg, Universal dynamics and deterministic switching of dissipative Kerr solitons in optical microresonators. *Nat. Phys.* **13**, 94–102 (2017).
28. H. Zhou, Y. Geng, W. Cui, S.-W. Huang, Q. Zhou, K. Qiu, and C. W. Wong, Soliton bursts and deterministic dissipative Kerr soliton generation in auxiliary-assisted microcavities, *Light: Science & Applications* **8**, 50 (2019).

29. X. Yi, Q. F. Yang, K. Y. Yang, and K. Vahala, Active capture and stabilization of temporal solitons in microresonators. *Opt. Lett.* **41**, 2037-2040 (2016).

30. B. Q. Shen, L. Chang, J. Q. Liu, H. M. Wang, Q.-F. Yang, C. Xiang, R. N. Wang, J. J. He, T. Y. Liu, W. Q. Xie, J. Guo, D. Kinghorn, L. Wu, Q.-X. Ji, T. J. Kippenberg, K. Vahala, and J. E. Bowers, Integrated turnkey soliton microcombs. *Nature* **582**, 365–369 (2020).

31. J. Q. Liu, E. Lucas, A. S. Raja, J. J. He, J. Riemensberger, R. N. Wang, M. Karpov, H. R. Guo, R. Bouchand, and T. J. Kippenberg, Photonic microwave generation in the X- and K-band using integrated soliton microcombs. *Nature Photon.* **14,** 486–491 (2020).

32. L. Chang, W. Q. Xie, H. W. Shu, Q.-F. Yang, B. Q. Shen, A. Boes, J. D. Peters, W. Jin, C. Xiang, S. T. Liu, G. Moille, S.-P. Yu, X. J. Wang, K. Srinivasan, S. B. Papp, K. Vahala and J. E. Bowers, Ultra-efficient frequency comb generation in AlGaAs-on-insulator microresonators. *Nature Commun.* **9**, 1869 (2018).

33. Z. Gong, A. Bruch, M. Shen, X. Guo, H Jung, L. Fan, X. W. Liu, L. Zhang, J. X. Wang, J. M. Li, J. C. Yan, and H. X. Tang, High-fidelity cavity soliton generation in crystalline AlN microring resonators. *Opt. Lett.* **43**, 4366–4369 (2018).

34. C. Xiang, J. Liu, J. Guo, L. Chang, R. N. Wang, W. Weng, J. Peters, W. Xie, Z. Zhang, J. Riemensberger, J. Selvidge, T. J. Kippenberg, and J. E. Bowers, Laser soliton microcombs heterogeneously integrated on silicon. *Science* **373**, 6550 (2021).

35. B. Stern, X. C. Ji, Y. Okawachi, A. L. Gaeta, and M. Lipson, Battery-operated integrated frequency comb generator. *Nature* **562**, 401-415 (2018).

36. A. S. Raja, A. S. Voloshin, H. R. Guo, S. E. Agafonova, J. Q. Liu, A. S. Gorodnitskiy, M. Karpov, N. G. Pavlov, E. Lucas, R. R. Galiev, A. E. Shitikov, J. D. Jost, M. L. Gorodetsky, and T. J. Kippenberg, Electrically pumped photonic integrated soliton microcomb. *Nat. Commun.* **10**, 680 (2019); correction **10**, 1623 (2019).

37. E. Nazemosadat, A. Fülöp, O. B. Helgason, P. H. Wang, Y. Xuan, D. E. Leaird, M. H. Qi, E. Silvestre, A. M. Weiner, and V. Torres-Company, Switching dynamics of dark-pulse Kerr frequency comb states in optical microresonators. *Phys. Rev. A* **103**, 013513 (2021).

38. M. H. Anderson, G. Lihachev, W. L. Weng, J.Q. Liu, and T. J. Kippenberg, Zero-dispersion Kerr solitons in optical microresonators. arXiv:2007.14507.

39. G. Lihachev, J. Q. Liu, W. L. Weng, L. Chang, J. Guo, J. J. He, R. N. Wang, M. H. Anderson, J. E. Bowers, and T. J. Kippenberg, Platicon microcomb generation using laser self-injection locking. arXiv:2103.07795.

40. H. W. Shu, L. Chang, C. H. Lao, B. T. Shen, W. Q. Xie, X. G. Zhang, M. Jin, Y. S. Tao, R. X. Chen, Z. H. Tao, S. H. Yu, Q.-F. Yang, X. J. Wang, and J. E. Bowers, Sub-milliwatt, widely-tunable coherent microcomb generation with feedback-free operation. arXiv:2112.08904.

41. S.-W. Huang, H. Liu, J. Yang, M. Yu, D.-L. Kwong, and C. W. Wong, Smooth and flat phase-locked Kerr frequency comb generation by higher order mode suppression. *Sci. Rep.* **6**, 26255 (2016).

42. C. Y. Bao, Y. Xuan, D. E. Leaird, S. Wabnitz, M. H. Qi, and A. M. Weiner, Spatial mode-interaction induced single soliton generation in microresonators. *Optica* **4**, 1011-1015 (2017).

43. W. Jin, Q.-F. Yang, L. Chang, B. Q. Shen, H. M. Wang, M. A. Leal, L. Wu, M. D. Gao, A. Feshali, M. Paniccia, K. J. Vahala, and J. E. Bowers, Hertz-linewidth semiconductor lasers using CMOS-ready ultrahigh-Q microresonators. *Nat. Photonics* **15**, 1–8 (2021).

44. H. Liu, S.-W. Huang, W. Wang, J. Yang, M. Yu, D.-L. Kwong, P. Colman, and C. W. Wong, Stimulated generation of deterministic platicon frequency microcombs. *Photonics Res.* **10**, 1877(2022).

45. Y. Li, S.-W. Huang, B. Li, H. Liu, J. Yang, A. K. Vinod, K. Wang, M. Yu, D.-L. Kwong, H. Wang, K. K.-Y. Wong, and C. W. Wong, Real-time transition dynamics and stability of chip-scale dispersion-managed frequency microcombs, *Light: Science & Applications* **9**, 52 (2020).

46. X. Ji, J. K. Kang, U. D. Dave, M. Corato-Zanarella, C. Joshi, A. L. Gaeta, and M. Lipson, Exploiting ultralow loss multimode waveguides for broadband frequency combs, *Laser & Photonics Rev.* **15**, 2000353 (2021).

47. K. Tamura, E. P. Ippen, H. A. Haus, and L. E. Nelson, 77-fs pulse generation from a stretched-pulse mode-locked all-fiber ring laser. *Opt. Lett.* **18**, 1080-1082 (1993).

48. C. Y. Bao and C. X. Yang, Stretched cavity soliton in dispersion-managed Kerr resonators. *Phys. Rev. A* **92**, 023802 (2015).

49. W. Wang, H. Liu, J. Yang, A. K. Vinod, J. Lim, Y.-S. Jang, H. Zhou, M. Yu, G.-Q. Lo, D.-L. Kwong, P. DeVore, J. Chou, and C. W. Wong, Mapping ultrafast timing jitter in dispersion-managed 89 GHz frequency microcombs via self-heterodyne linear interferometry. arXiv:2108.01177.

50. N. L. B. Sayson, K. E. Webb, S. Coen, M. Erkintalo, and S. G. Murdoch, Widely tunable optical parametric oscillation in a Kerr microresonator. *Opt. Lett.* **42**, 5190-5193 (2017).
51. C. J. Bao, H. Taheri, L. Zhang, A. Matsko, Y. Yan, P. C. Liao, L. Maleki, and A. E. Willner, High-order dispersion in Kerr comb oscillators. *J. Opt. Soc. Am. B.* **34**, 715-725 (2017).
52. D. J. Jones, Y. Chen, H. A. Haus, and E. P. Ippen, Resonant sideband generation in stretched-pulse fiber lasers. *Opt. Lett.* **23**, 1535-1537 (1998).
53. X. Yi, Q. F. Yang, X. Zhang, K. Y. Yang, X. Li, and K. Vahala, Single-mode dispersive waves and soliton microcomb dynamics. *Nat. Commun.* **8**, 14869 (2017).
54. Q.-F. Yang, Q.-X. Ji, L. Wu, B. Shen, H. Wang, C. Bao, Z. Yuan, and K. Vahala, Dispersive-wave induced noise limits in miniature soliton microwave sources. *Nat. Commun.* **12**, 1442 (2021).